\documentclass{jfm}
\usepackage{upgreek}
\usepackage{esint}
\usepackage{tabularx}
\newcommand{\be}{\begin{equation}}
\newcommand{\ee}{\end{equation}} 
\newcommand{\bea}{\begin{eqnarray}}
\newcommand{\eea}{\end{eqnarray}}
\usepackage{todonotes}
\begin{document}
\newtheorem{lemma}{Lemma}
\newtheorem{corollary}{Corollary}

\shorttitle{Revisiting Bolgiano-Obukhov scaling for moderately stably stratified turbulence } 
\shortauthor{Shadab Alam, Anirban Guha and Mahendra K. Verma} 

\title{Revisiting Bolgiano-Obukhov scaling for moderately stably stratified turbulence}

\author
 {
 Shadab Alam\aff{1}
  Anirban Guha\aff{1},\aff{2}
  \corresp{\email{anirbanguha.ubc@gmail.com}},
 and
Mahendra K. Verma\aff{3}  
  }

\affiliation
{
\aff{1}
Department of Mechanical Engineering, Indian Institute of Technology Kanpur, Kanpur 208016.
\aff{2} Institute of Coastal Research, Helmholtz-Zentrum Geesthacht, Geesthacht 21502, Germany.
\aff{3}
Department of Physics, Indian Institute of Technology Kanpur, Kanpur 208016.
}

\maketitle

\begin{abstract}
According to the  celebrated Bolgiano--Obukhov \citep{Bolgiano_1959,Obukhov_1959} phenomenology for moderately stably stratified turbulence, the energy spectrum in the inertial range shows a dual scaling; the kinetic energy follows (i)  $\sim k^{-11/5}$ for $k < k_B$, and (ii)  $\sim k^{-5/3}$ for $k > k_B$, where $k_B$ is Bolgiano wavenumber. The $k^{-5/3}$ scaling akin to passive scalar turbulence is a direct consequence of the assumption that  buoyancy is insignificant  for $k>k_B$. We revisit this assumption, and using constancy of kinetic and potential energy fluxes and simple theoretical analysis, we find that $ k^{-5/3}$ spectrum is absent.  This is because the velocity field at small scales is too weak to establish a constant kinetic energy flux as in passive scalar turbulence.  A quantitative condition for the existence of the second regime is also derived in the paper.   
\end{abstract}

\section{Introduction}\label{sec:introduction}
Stable density stratification is  commonly observed  in oceans and the nocturnal atmosphere \citep{Sagaut:book,Turner:book,Davidson:book:TurbulenceRotating,
Maffioli_Davidson_2016}.
Both  atmospheric and oceanic flow can often be turbulent; such turbulence, commonly known as ``stably stratified turbulence'' (SST), is different from the classical ``Kolmogorov turbulence'', which is applicable to homogeneous and isotropic hydrodynamic turbulence.  

Stably stratified turbulence is quite complex, and there are many unresolved issues in this field~\citep{lindborg_2006,Lindborg:JFM2007,Brethouwer:JFM2007, Davidson:book:TurbulenceRotating,Rosenberg_pof2015, Verma_BDF_2018}.  One of the important parameters here is Froude number 
\be
Fr \equiv \frac{U}{NL},
\ee
where $U$ and $L$  are  the large scale velocity and length scale respectively, and $N$ is the {\em Brunt-V\"{a}is\"{a}l\"{a}} frequency (defined in \S 2)~\citep{Davidson:book:TurbulenceRotating}.  
A related parameter is the Richardson number 
\be
Ri \equiv \frac{N^2}{(\partial u \slash \partial z)^2},
\ee
 which is the ratio of buoyancy and flow shear.  The approximation  $\partial u \slash \partial z \sim U/L$ yields $Ri \approx Fr^{-2}$~(\citealt{Rosenberg_pof2015};~\citealt*{ Maffioli_2016}). The degree of turbulence is quantified by the Reynolds number: $Re \equiv U L/\nu$, where $\nu$ is the kinematic viscosity of the fluid.

Based on the above parameters, stably stratified turbulent flows can be classified into three broad regimes:
\begin{itemize}
\item $Re\gg 1$ and $Fr\gg 1$ (turbulent SST with weak buoyancy): In this regime, strong nonlinearity (${\bf u \cdot \nabla u}$)  in comparison to buoyancy yields scaling similar to passive scalar turbulence. Hence, both kinetic energy spectrum, $E_u(k)$, and potential energy spectrum, $E_b(k)$, follow  Kolmogorov spectrum and scale as $\sim k^{-5/3}$, where $k$ denotes wavenumber. 

 \item $Re\gg 1$ and $Fr\ll 1$ (turbulent SST with strong buoyancy): Here, buoyancy is much stronger than the nonlinearity.  The flow is strongly anisotropic with strong horizontal velocity compared to the vertical velocity~\citep{Davidson:book:TurbulenceRotating,Vallis_1997,
lindborg_2006,Lindborg:JFM2007,Brethouwer:JFM2007}.  The flow of the terrestrial atmosphere is strongly stratified with typical $Fr \sim 0.01$~\citep{Waite_2011}.The physics of this regime is quite complex, and it is still being debated.

\item $Re\gg 1$  and $Fr\approx 1$ (turbulent SST with moderate buoyancy): Here,  buoyancy and nonlinearity are of comparable strength.  \citet{Bolgiano_1959} and \citet{Obukhov_1959} constructed a model for this regime by arguing that  buoyancy force converts kinetic energy  into  potential energy.   They argued for a dual scaling with transition occurring at Bolgiano wavenumber $k_B$.  For $k < k_B$, the kinetic energy flux $\Pi_u(k)$ decreases   as $\sim k^{-4/5}$, but    the potential energy flux $\Pi_b(k)$ is constant.  Here,
$E_u(k) \sim k^{-11/5}$, and $E_b(k) \sim k^{-7/5}$.  For $k > k_B$, buoyancy is expected to be weak, and hence the scaling is similar to that for passive scalar.  We denote the above model as BO phenomenology. 
\end{itemize}

In this paper we focus on the third regime---moderately stratified turbulence.  For this, computational studies by~\citet{Waite_bartello_2004}, and~\citet*{kumar_2014} show that the flow remains approximately isotropic. Furthermore, direct numerical simulation results of \citet{kimura_herring_1996} and \citet{kumar_2014}, shell-model results of \citet{kumar_2015}, and global energy balance analysis of \citet{BHATTACHARJEE_2015} have unequivocally shown that the kinetic energy spectrum indeed scales as $\sim k^{-11/5}$ in a wavenumber band.  However, we are not aware of any numerical or experimental work that convincingly demonstrate the dual scaling for such flows.   

Several researchers have reported Bolgiano-Obukhov scaling for  turbulent thermal convection, Rayleigh-Taylor turbulence, and unstably stratified flows. Note however that  \citet*{Verma_2017} showed that Bolgiano-Obukhov scaling is not applicable to such flows in three dimensions; instead they follow Kolmogorov-like turbulence phenomenology ($E_u(k) \sim k^{-5/3}$). Yet, in two dimensional  turbulent thermal convection and Rayleigh-Taylor turbulence exhibit Bolgiano-Obukhov scaling, as demonstrated by \citet{Boffetta:JFM2012} and \citet{Boffetta:ARFM2017}. 
\citet{Verma_2017} and \citet{Verma_BDF_2018}   argued that the above phenomena is due to the inverse cascade of kinetic energy.

 The flow in the deep oceans are moderately stratified with $Fr \sim 1$~\citep{Petrolo_Woods_2019}.  The atmosphere of some other planets could yield a wide range of $Fr$; hence, a clear understanding of SST with moderate buoyancy is essential.
The goal of this paper is to revisit turbulent SST with moderate buoyancy and critically examine the validity of dual scaling of the BO phenomenology. We start with the constancy of total energy flux (kinetic + potential) and demonstrate that for large wavenumbers, the velocity field becomes weak; hence, the assumption that  buoyancy becomes weak at large wavenumbers leading to $k^{-5/3}$ spectra is improbable.  We observe that $E_u(k) \sim k^{-11/5}$ for $k > 1/L$, where $L$ is the system size, with no crossover to $k^{-5/3}$ spectra. As an aside, we recover $E_u(k) = k^{-5/3}$ for $k < 1/L$, which may be possible in systems with large aspect ratio.  Thus, we provide a revision of the celebrated Bolgiano and Obukhov phenomenology.

The outline of the paper is as follows:  The equations  governing SST  are introduced in \S \ref{sec:governing}.  The BO phenomenology is described in \S \ref{sec:bo_pheno}. In \S \ref{sec:numerical solution} and \S  \ref{sec:asymptotic analysis}, respectively, numerical solution  and asymptotic analysis of the equation for the total energy flux (a fifth order equation)  are presented. We conclude in  \S \ref{sec:conclusion}.

\section{Governing Equations}\label{sec:governing} 
The governing Navier-Stokes equations for  stably stratified flows (density stratification in the vertical ($z$) direction) under the Boussinesq  approximation are~\citep{Davidson:book:Turbulence,lindborg_2006,Davidson:book:TurbulenceRotating,Verma_BDF_2018}
\begin{subequations}
\bea
\frac{\partial{\mathbf{u}}}{\partial{t}}+ (\mathbf{u}\cdot \nabla)\mathbf{u} & = &  -\frac{1}{\rho_m}\nabla\sigma - N b \hat{z} + \nu \nabla^{2}\mathbf{u} +{\bf F}_u,  
\label{eq:u2_SS} \\
\frac{\partial{b}}{\partial{t}}+(\mathbf{u}\cdot\nabla)b & = & N  u_{z} + \kappa \nabla^{2}b, 
\label{eq:b2_SS} \\
\nabla \cdot \mathbf{u}& = & 0,
\eea
\end{subequations}
 where ${\bf u}=(u_x,\,u_y,\,u_z)$ and $\sigma$ are respectively the velocity and the  pressure fields;   $\nu$ and $\kappa$ are respectively  the kinematic viscosity and diffusivity of the density fluctuation; $\rho_m$ is the  mean density; ${\bf F}_u$ is the external force (in addition to the buoyancy); and $b$ is the density fluctuation  in  velocity unit, which is achieved by the following transformation \citep{lindborg_2006,Davidson:book:TurbulenceRotating,Rosenberg_pof2015}:
\index{Density!fluctuation $b$}
\be
b  = \frac{g}{N}\frac{\rho}{\rho_m}, 
\label{eq:b_def}
 \ee
where $g$ is the acceleration due to gravity, and $\rho$ is the density fluctuation. The quantity
\be
 N = \sqrt{\frac{g}{\rho_m} \left| \frac{d \bar{\rho}}{d z} \right|}
 \ee
is the {\em Brunt-V\"{a}is\"{a}l\"{a} frequency}.   Note that $-Nb$ is   buoyancy. 


It is convenient to describe the flow behaviour in the Fourier space since it captures the scale-by-scale energy transfer and interactions. The following one-dimensional kinetic spectrum, $E_u(k)$, and the potential energy spectrum, $E_b(k)$, that are sum of respective energy of all the modes of a shell of thickness $dk$, are introduced. 
\bea
E_u(k,t)dk & = & \sum_{k < |{\bf k'}| \le k + dk} \frac{1}{2} |{\bf u(k'},t)|^2, \\
E_b(k,t)dk & = & \sum_{k < |{\bf k'}| \le k + dk} \frac{1}{2} |b{\bf(k'},t)|^2.
\label{eq:SST_E_ub_k}
\eea
Note that $E_u(k)$ and $E_b(k)$ are averaged over polar angles, hence they do not capture the anisotropic effects.  Ring spectrum, proposed by~\citet{ Teaca:PRE2009} and~\citet{Nath:PRF2016}, captures the angular dependent spectra.  

Henceforth, the explicit time dependence in $E_u$ and $E_b$ are suppressed for brevity. The nonlinear energy transfers across modes are quantified using energy fluxes or energy cascade rates.  The kinetic (potential) energy flux, $\Pi_{u(b)}(k_0)$, for a wavenumber sphere of radius $k_0$ is the total kinetic (potential) energy leaving the said sphere due to nonlinear interactions. These fluxes are computed using the  following formulas \citep{Dar:PD2001,Verma:PR2004,Verma_BDF_2018}:
\begin{subequations}
\bea
\Pi_{u}(k_0) & = & \sum_{|{\bf k}| > k_0}   \sum_{|{\bf p}| \leq k_0}
 \Im \left[  {\bf  \{  k \cdot u(q) \} \{ u({\bf p}) \cdot u^*({\bf k}) \} }  \right] ,\\
\Pi_{b}(k_0) & = & \sum_{|{\bf k}|>k_0}   \sum_{|{\bf p}| \leq k_0} 
\Im \left[  {\bf  \{  k \cdot u(q) \} } \{ b({\bf p})  b^*({\bf k}) \}  \right],
\eea
\end{subequations}
where ${\bf k= p+q}$. 



\begin{figure}
\begin{center}
\includegraphics[scale = 0.5]{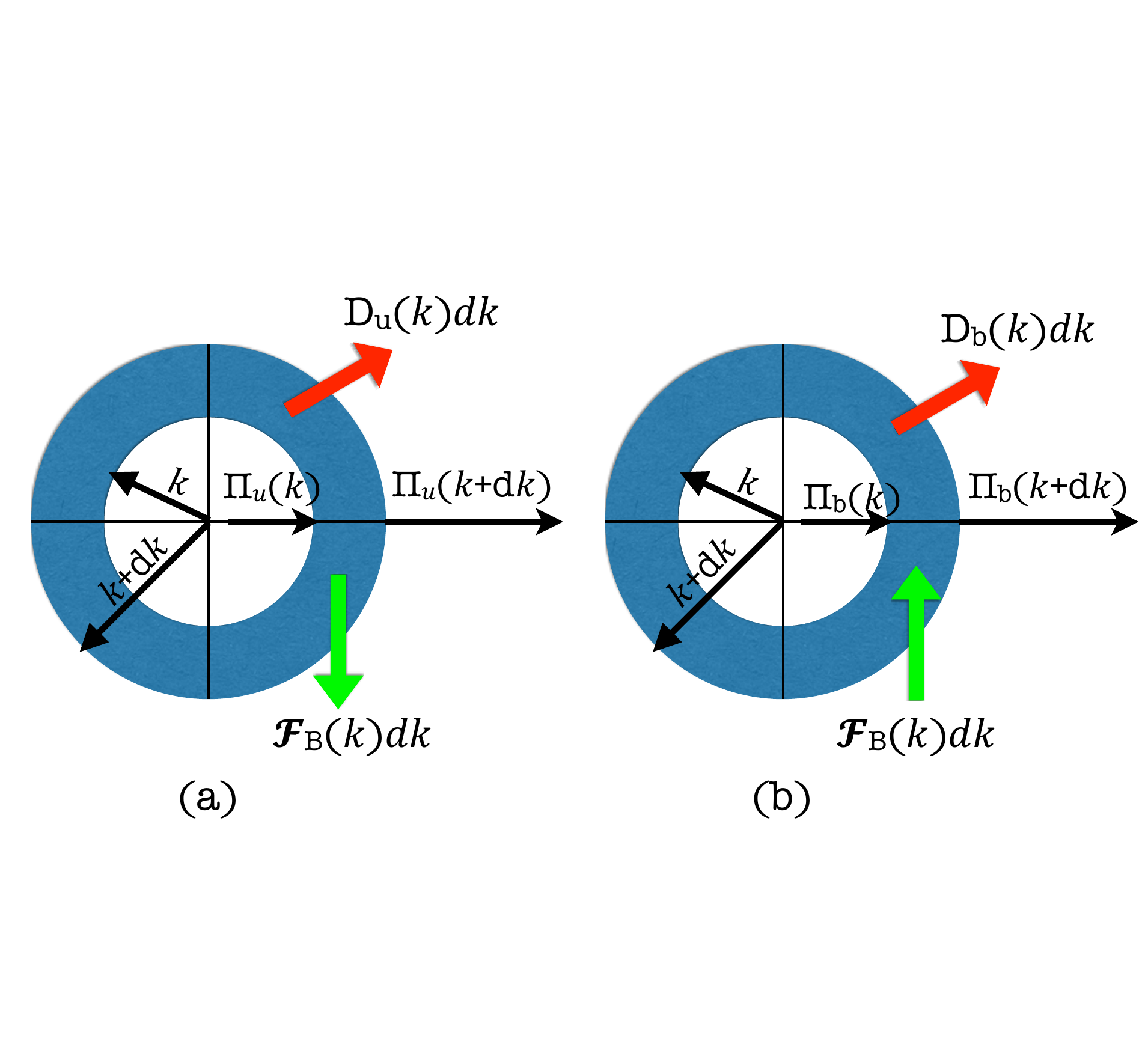}
\end{center}
\vspace*{-50pt}
\caption{(Color online) (a) The kinetic energy contents of a wavenumber shell changes due to the kinetic energy flux difference $\Pi_u(k+dk)-\Pi_u(k)$, energy removal rate by buoyancy $\mathcal{F}_B(k)dk$, and viscous dissipation rate $D_u(k)dk$. (b) The potential energy changes due to potential energy flux difference $\Pi_b(k+dk)-\Pi_b(k)$, energy supply rate by buoyancy $\mathcal{F}_B(k)dk$,  and  dissipation rate $D_b(k)dk$.}
\label{fig:Pi_shell}
\end{figure}

The dynamical equations for modal kinetic energy ($E_u({\bf k}) = \frac{1}{2} |{\bf u(k})|^2 $) and potential energy ($E_b({\bf k}) = \frac{1}{2} |b{\bf (k})|^2 $) respectively can be derived from (\ref{eq:u2_SS}) and  (\ref{eq:b2_SS}), and are as follows~\citep{Davidson:book:TurbulenceRotating, Verma_BDF_2018}:

\begin{subequations}
\bea
\frac{d}{dt}E_u({\bf k}) &=& T_u({\bf k})+\mathcal{F}_B({\bf k})+\mathcal{F}_{ext}({\bf k})-D_u({\bf k}),
\label{eq:E_u_SS}\\ 
\frac{d}{dt}E_b({\bf k}) &=& T_b({\bf k})-\mathcal{F}_B({\bf k})-D_b({\bf k}).
\label{eq:E_b_SS} 
\eea
\end{subequations}
Here $T_{u(b)}({\bf k})$  and  $D_{u(b)}({\bf k})$ are respectively the nonlinear kinetic (potential) energy transfer rate and dissipation rate, while  $\mathcal{F}_{B}$ and $\mathcal{F}_{ext}$  denote  the energy feed rate by the buoyancy and external force respectively. These quantities are defined as follows \citep{Verma_2017,Verma_BDF_2018}:
\begin{subequations}
\bea
T_{u}({\bf k}) & = &   \sum_{{\bf p}}
 \Im \left[  {\bf  \{  k \cdot u(q) \} \{ u({\bf p}) \cdot u^*({\bf k}) \} }  \right] ,\\
T_{b}({\bf k}) & = &   \sum_{{\bf p}} 
\Im \left[  {\bf  \{  k \cdot u(q) \} } \{ b({\bf p})  b^*({\bf k}) \}  \right],\\
\mathcal{F}_B({\bf k}) & = & -  N  \Re \left[b({\bf k}) u_z^*({\bf k})\right],\\
\mathcal{F}_{ext}({\bf k}) & = & \Re \left[{\bf F}_u({\bf k}) \cdot {\bf u}^*({\bf k})\right],\\
D_u({\bf k}) & = & 2\nu k^2E_u({\bf k}),\\
D_b({\bf k}) & = & 2\kappa k^2E_b({\bf k}),
\eea
\end{subequations}
where {\bf k = p + q}. The kinetic and potential energy fluxes are related to nonlinear energy transfer terms as
\renewcommand{\theequation}
{\arabic{section}.\arabic{equation}a,b}
\bea
\Pi_u({\bf k_0}) = - \sum_{|{\bf k}| \leq k_0}T_u({\bf k}); \ \ \ \ \ \ \ \ 
\Pi_b({\bf k_0}) = - \sum_{|{\bf k}| \leq k_0}T_b({\bf k}).
\eea
\renewcommand{\theequation}{\arabic{section}.\arabic{equation}}
We write  (\ref{eq:E_u_SS}) and (\ref{eq:E_b_SS}) for the spheres of radii $k$ and $k+dk$ and take their difference that yields  
\begin{subequations}
\bea
\frac{d}{dt}\sum_{k <  |{\bf k'}| \le k + dk}E_u({\bf k'}) &=& \sum_{k <  |{\bf k'}| \le k + dk}T_u({\bf k'})+\mathcal{F}_B({\bf k'})+ \mathcal{F}_{ext}({\bf k'}) - D_u({\bf k'}),
\label{eq:E_u_SS_sum}\\ 
\frac{d}{dt}\sum_{k < |{\bf k'}| \le k + dk}E_b({\bf k'}) &=& \sum_{k <  |{\bf k'}| \le k + dk}T_b({\bf k'})-\mathcal{F}_B({\bf k'})- D_b({\bf k'}),
\label{eq:E_b_SS_sum} 
\eea
\end{subequations}
where
\begin{subequations}
\bea
 \sum_{k <  |{\bf k'}| \le k + dk}T_u({\bf k'}) &=& -\Pi_u(k+dk) + \Pi_u(k),\\
  \sum_{k <  |{\bf k'}| \le k + dk}T_b({\bf k'}) &=& -\Pi_b(k+dk) + \Pi_b(k).
\eea
\end{subequations}
Now taking the limit $dk \rightarrow 0$ yields
\begin{subequations}
\bea
\frac{d}{dt}E_u(k) &=& -\frac{d}{dk}\Pi_u(k)+\mathcal{F}_B(k)+\mathcal{F}_{ext}(k)-D_u(k),
\label{eq:Eub_SS}\\ 
\frac{d}{dt}E_b(k) &=& -\frac{d}{dk}\Pi_b(k)-\mathcal{F}_B(k)-D_b(k),
\label{eq:E_ub_SS} 
\eea
\end{subequations}
where
\begin{subequations}
\bea
\mathcal{F}_B(k)dk & = & -   \sum_{k <  |{\bf k'}| \le k + dk}N  \Re \left[b({\bf k'}) u_z^*({\bf k'})\right],\\
\mathcal{F}_{ext}(k)dk & = &  \sum_{k <  |{\bf k'}| \le k + dk}\Re \left[{\bf F}_u({\bf k'}) \cdot {\bf u}^*({\bf k'})\right],\\
D_u(k)dk & = & 2\nu  \sum_{k <  |{\bf k'}| \le k + dk}k'^2E_u({\bf k'}),\\
D_b(k)dk & = & 2\kappa \sum_{k <  |{\bf k'}| \le k + dk} k'^2E_b({\bf k'}).
\eea
\end{subequations}
The above energetics is illustrated in figure \ref{fig:Pi_shell}.


Let us consider a statistically steady state ($\partial  /\partial t \rightarrow 0$). In the  inertial range, $\mathcal{F}_{ext} = 0$, and the dissipative effects are  negligible, i.e.\ $D_u \rightarrow 0$  and $D_b \rightarrow 0$.  Hence the  equations for the kinetic and potential energies simplify to 
\begin{subequations}
\bea
  \frac{d}{dk} \Pi_u(k)  &=&  \mathcal{F}_B(k),  
   \label{eq:SST_energetics_steady_dPiu} \\
  \frac{d}{dk} \Pi_b(k)  &=& - \mathcal{F}_B(k).
  \label{eq:SST_energetics_steady_dPirho}
  \eea
\end{subequations}
The sum of (\ref{eq:SST_energetics_steady_dPiu}) and (\ref{eq:SST_energetics_steady_dPirho}) yield
\be
\Pi_u(k) + \Pi_b(k) = \Pi = \mathrm{constant} .
\label{eq:SST_total_flux}
\ee   
Hence the total energy flux is constant in the inertial range.  We will employ  (\ref{eq:SST_total_flux}) in later part of the paper.

Based on energetics arguments, it has been shown that the energy injection rate by buoyancy, $\mathcal{F}_B$, is negative.  Hence $\Pi_u(k)$ decreases with $k$~\citep{kumar_2014, Verma_BDF_2018, Verma_2019}.~\citet{Verma_2019} showed that in the linear regime,  gravity waves facilitate periodic exchange of kinetic and potential energies, hence $\mathcal{F}_B =0$.  Therefore,  a nondissipative gravity wave represents a neutral state.  Since the system is stable, the nonlinearity makes $\mathcal{F}_B$ negative.  If $\mathcal{F}_B >0$, according to the integral form of (\ref{eq:E_u_SS}), the kinetic energy would grow in time, thus making the flow unstable.  Hence, $\mathcal{F}_B <0$.  In addition,~\cite{kumar_2014} and \cite{Verma_2019} go on to argue that $\mathcal{F}_B(k) <0$. The above features have been verified numerically by \citet{kumar_2014} and \cite{Verma_2017}.

When we substitute negative $\mathcal{F}_B(k)$ in  (\ref{eq:SST_energetics_steady_dPiu}, \ref{eq:SST_energetics_steady_dPirho}), we deduce that $\Pi_u(k) $ decreases with $k$, while $\Pi_b(k) $ increases with $k$.  These features play an important role in the models of  \citet{Bolgiano_1959} and \citet{Obukhov_1959}.

The  equations described in this section applies to all the three regimes.  In the following two sections we will focus on phenomenology of moderately stratified turbulence.

\section{ The Bolgiano--Obukhov phenomenology for moderately stably stratified turbulence} 
\label{sec:bo_pheno}

 \citet{Bolgiano_1959} and \citet{Obukhov_1959} constructed a phenomenology for moderately stratified turbulence, which we refer to as BO phenomenology. In this regime, the flow is nearly isotropic.~\citet{kumar_2014} showed that for $Fr \gtrsim 1$, the anisotropic ratio $E_\perp/2E_\parallel \approx 1$, where $E_\perp = (u_x^2+u_y^2)/2$ and $E_\parallel = u_z^2/2$. \citet{Waite_bartello_2004} also showed that the flow is approximately isotropic for $Fr = 1.3$, and  anisotropy of stratification starts to become visible for $Fr \le 0.21$. It has been conjectured that isotropy is also present in the inertial range of moderately SST.

According to the BO phenomenology, a force balance between the nonlinear term and  buoyancy in (\ref{eq:u2_SS}) yields 
\be
k u_k^2 = N b_k,
\label{eq:force_balance_b}
\ee
where $u_k$ and $b_k$ are respectively the velocity and density fluctuations  at wavenumber $k$. In addition, the BO phenomenology assumes that in the inertial range, $\Pi_b(k)\approx \mathrm{constant}$, and it equals the dissipation rate  of the potential energy ($\epsilon_b$):
\be
\Pi_b(k)  =  k b_k^2 u_k = \epsilon_b.
\label{eq:SST_const_Pi_rho}
\ee
Equations (\ref{eq:force_balance_b}) and  (\ref{eq:SST_const_Pi_rho}) yield the following relations:
  \begin{subequations}
 \begin{eqnarray}
E_u(k) & =  & \frac{u_k^2}{k} = c_1 \epsilon_b^{2/5}  N^{4/5} k^{-11/5}, \label{eq:Eu_2} \\
E_b(k) & = & \frac{b_k^2}{k} = c_2 \epsilon_b^{4/5}  N^{-2/5} k^{-7/5}, \label{eq:Eb} \\
\Pi_u(k) & = & k u_k^3 = c_3  \epsilon_b^{3/5} N^{6/5} k^{-4/5},  \label{eq:pi_u} \\
\Pi_b(k) & = &  \epsilon_b. \label{eq:pi_b} 
\end{eqnarray} 
  \end{subequations}

Bolgiano and Obukhov argued that the above-mentioned behaviour of the inertial range is  true only for  lower wavenumbers ($k < k_B$, where $k_B$ will be defined below). For $k > k_B$ of the inertial range, the buoyancy effects are weak and hence cannot balance the inertial term (which is balanced by the pressure gradient). Hence in this region, the  scaling of  passive scalar (i.e.\ Kolmogorov) turbulence should be valid. The energy and flux relations obtained here are:
\begin{subequations}
\bea
E_u(k) & =  & K_{Ko}  \epsilon_u^{2/3}k^{-5/3}, \label{eq:Eu_KO2} \\
E_b(k) & =  & K_\mathrm{OC} \epsilon_u^{-1/3}\epsilon_b k^{-5/3}, \label{eq:Eb_KO} \\
\Pi_u(k) & = &  \epsilon_u,  \label{eq:pi_u_KO} \\
\Pi_b(k) & = &  \epsilon_b, \label{eq:pi_b_KO}
\end{eqnarray}
\end{subequations}
where $\epsilon_u$ is the viscous dissipation rate; and $K_\mathrm{Ko}, K_\mathrm{OC}$ are Kolmogorov's and Obukhov-Corrsin's constants.   It is important to keep in mind that the viscous dissipation and thermal dissipation play a critical role in turbulence.  They set up the fluxes, $\Pi_u$ and $\Pi_b$,  even though they are not very active in the inertial range. 

\begin{figure}
\begin{center}
\includegraphics[scale = 0.6]{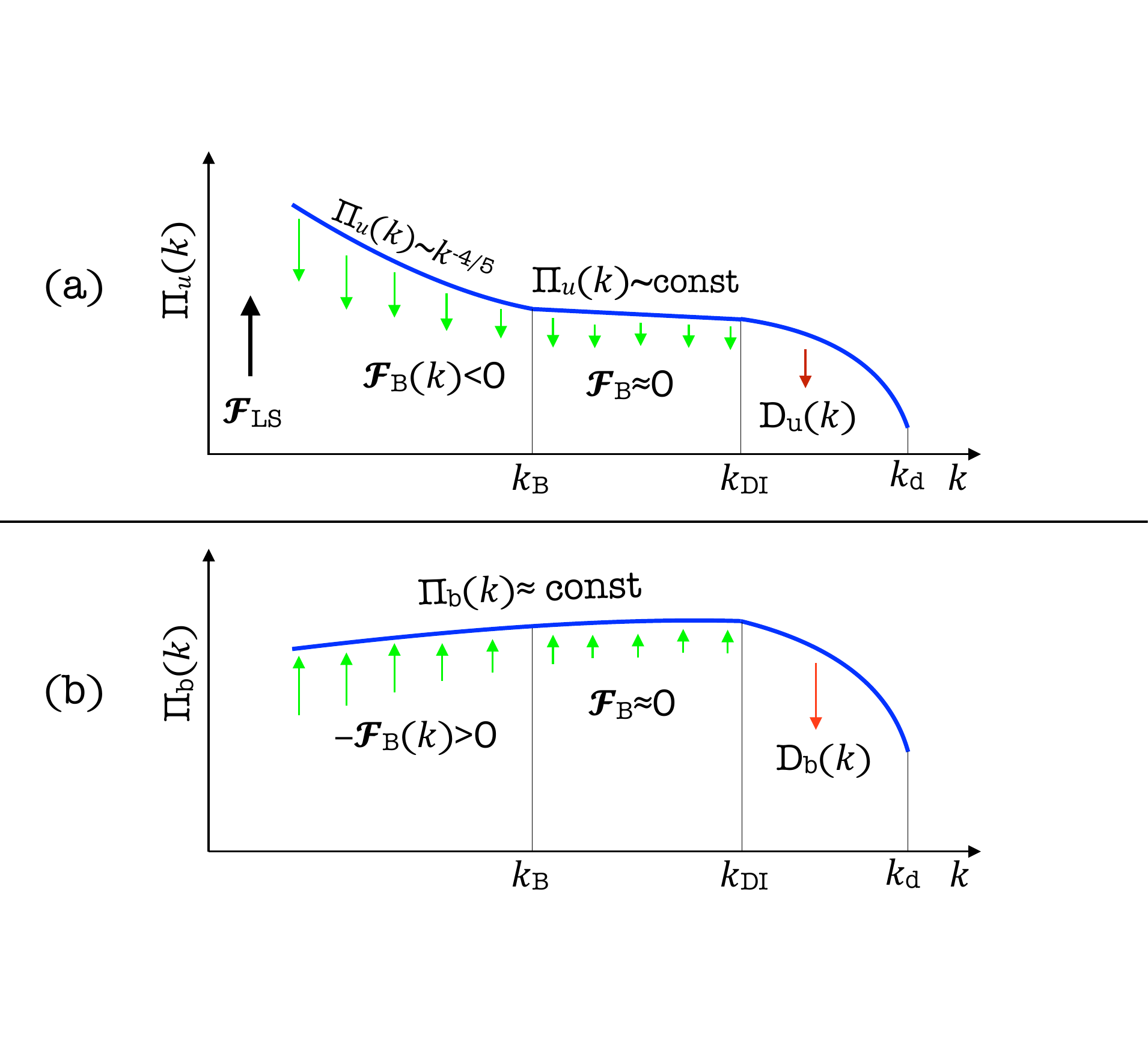}
\end{center}
\vspace*{-35pt}
\caption{(Color online) Schematic diagram of moderately stably stratified turbulence according to the Bolgiano-Obukhov phenomenology. (a) Kinetic energy flux and (b) potential energy flux. Transition from inertial regime to dissipation regime occur at wavenumber $k_{DI}$. Here $k_d$ is Kolmogorov wavenumber, and $k_d \gg k_{DI}$.}
\label{fig:Pi_schematic_sst_bo}
\end{figure}

The behavioural transition from one regime to another  occurs near the Bolgiano wavenumber $k_B$, which is obtained by matching $\Pi_u(k) $ in the two regimes:
\be 
k_B \approx N^{3/2} \epsilon_u^{-5/4} \epsilon_b^{3/4}.
\label{eq:kB}
\ee
The nature of kinetic and potential energy fluxes, as well as dual scaling of moderately stably stratified turbulence as predicted by Bolgiano and Obukhov are illustrated  in  figure \ref{fig:Pi_schematic_sst_bo}. We also remark that $\Pi_u(k)$ decreases rapidly as $k^{-4/5}$ and then it tapers of to $\epsilon_u$. However,  $\Pi_b \approx \epsilon_b \approx \Pi$ (see Eq.~(\ref{eq:SST_total_flux})).  Hence, $\epsilon_b \gg \epsilon_u$.

In addition to $k_B$, another important length referred to in SST is ``Ozmidov length", which is defined as
\be
L_O \equiv \sqrt{\frac{\epsilon_u}{N^3}}.
\label{eq:LO}
\ee
The corresponding wavenumber $k_O =1/L_O$.  At $L_O$, the time scales of gravity waves and local eddies match, i.e., $l/u_l \approx 1/N$.    Using a numerical simulation,~\citet{Waite_bartello_2004,Waite_bartello_2006} computed $L_O$  for $Fr = 1.3$ and reported that $L_O$ is approximately $1/31$ of the system size.

For moderately stratified flows, $\Pi_u(k)$ varies with $k$, hence it is not obvious whether we should substitute $\epsilon_u = \Pi_u(k)$ of (\ref{eq:pi_u}),  or $\epsilon_u$ of  (\ref{eq:pi_u_KO}).  In any case, it is interesting to compare $k_O$ with $k_B$.  Using  (\ref{eq:kB}, \ref{eq:LO}) we obtain
\be
\frac{k_B}{k_O}  = \epsilon_u^{-5/4+1/2}  \epsilon_b^{3/4} 
\sim \left( \frac{\epsilon_b}{\epsilon_u} \right)^{3/4}
\ee
Since $\epsilon_b \gg \epsilon_u$ for SST, we expect that $k_B \gg k_O$.

In the next section, we describe certain critical deficiencies of  the BO phenomenology.


\section{Revision of Bolgiano--Obukhov phenomenology for moderately stably stratified turbulence} \label{sec:bo revisited}

A crucial assumption made in the BO phenomenology is that  $ \Pi_b(k) \approx \mathrm{constant}$  in the inertial range (refer to (\ref{eq:SST_const_Pi_rho})).   This assumption needs a closer examination.  A more rigorous approach would be to start with the constancy of total energy flux  (equation (\ref{eq:SST_total_flux})) that follows from the conservation of total energy (kinetic + potential) in the inviscid limit.

We start with (\ref{eq:SST_total_flux}), and equate it  to the total dissipation rate $\epsilon$. That is,
\be
\Pi_u(k)  + \Pi_b(k) = k u_k^3 + k b_k^2 u_k  = \epsilon.
\label{eq:flux_sum2}
\ee
In the above equation we eliminate $b_k$ using (\ref{eq:force_balance_b}) that yields the following fifth-order polynomial in $u_k$:
\be
k u_k^3  +  \frac{k^3 u_k^5}{N^2} = \epsilon.
\label{eq:u_k_fifth_order}
\ee
There is no analytical solution for a fifth order algebraic polynomial.  Therefore, we employ numerical solution  and asymptotic analysis to solve the above equation.  These two results are consistent with each other. 

\subsection{Numerical Solution} \label{sec:numerical solution}
\begin{figure}
\begin{center}
 \begin{tabular}{cc}
 \includegraphics[scale = 0.38]{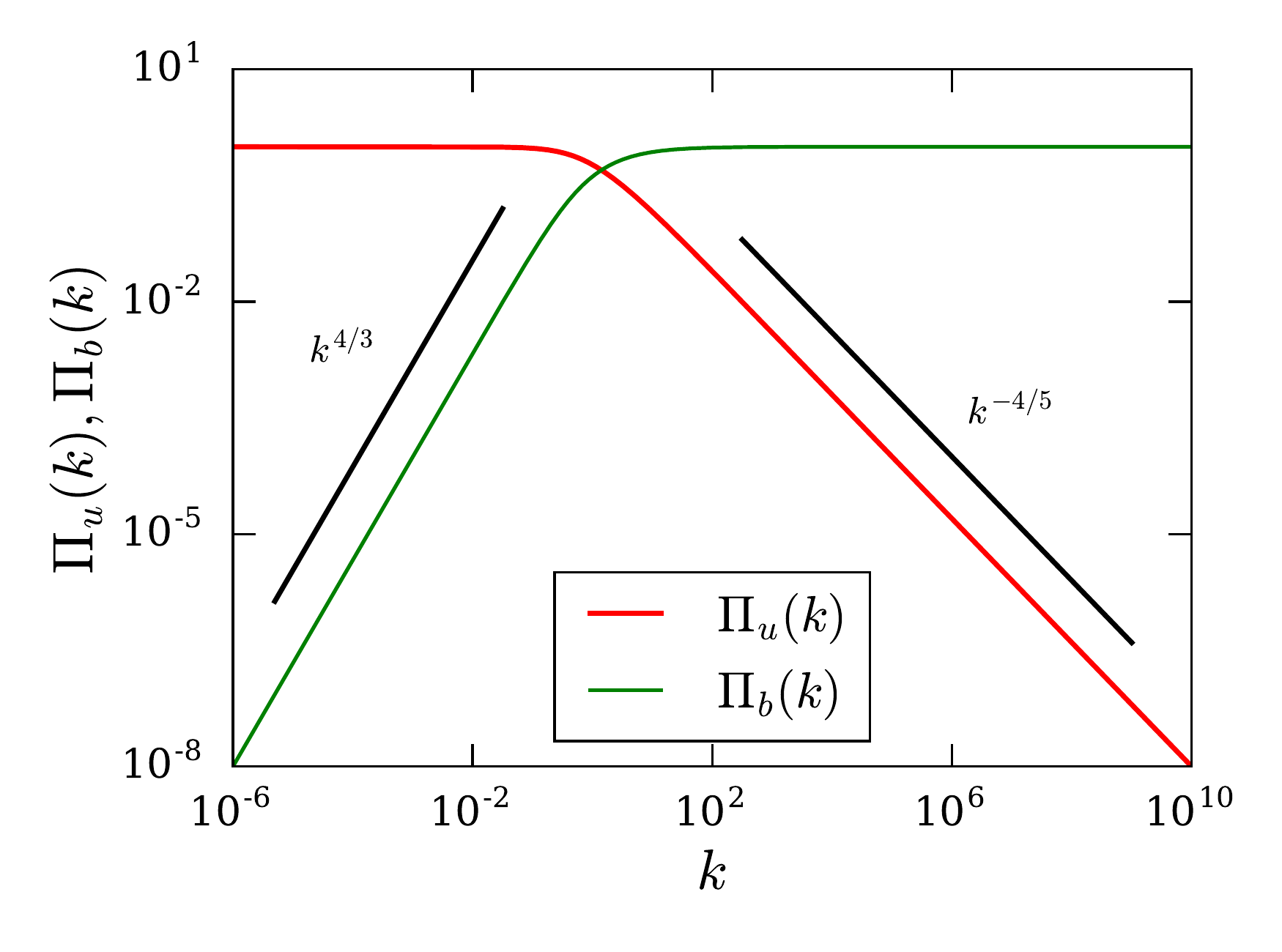}
&  \includegraphics[scale = 0.38]{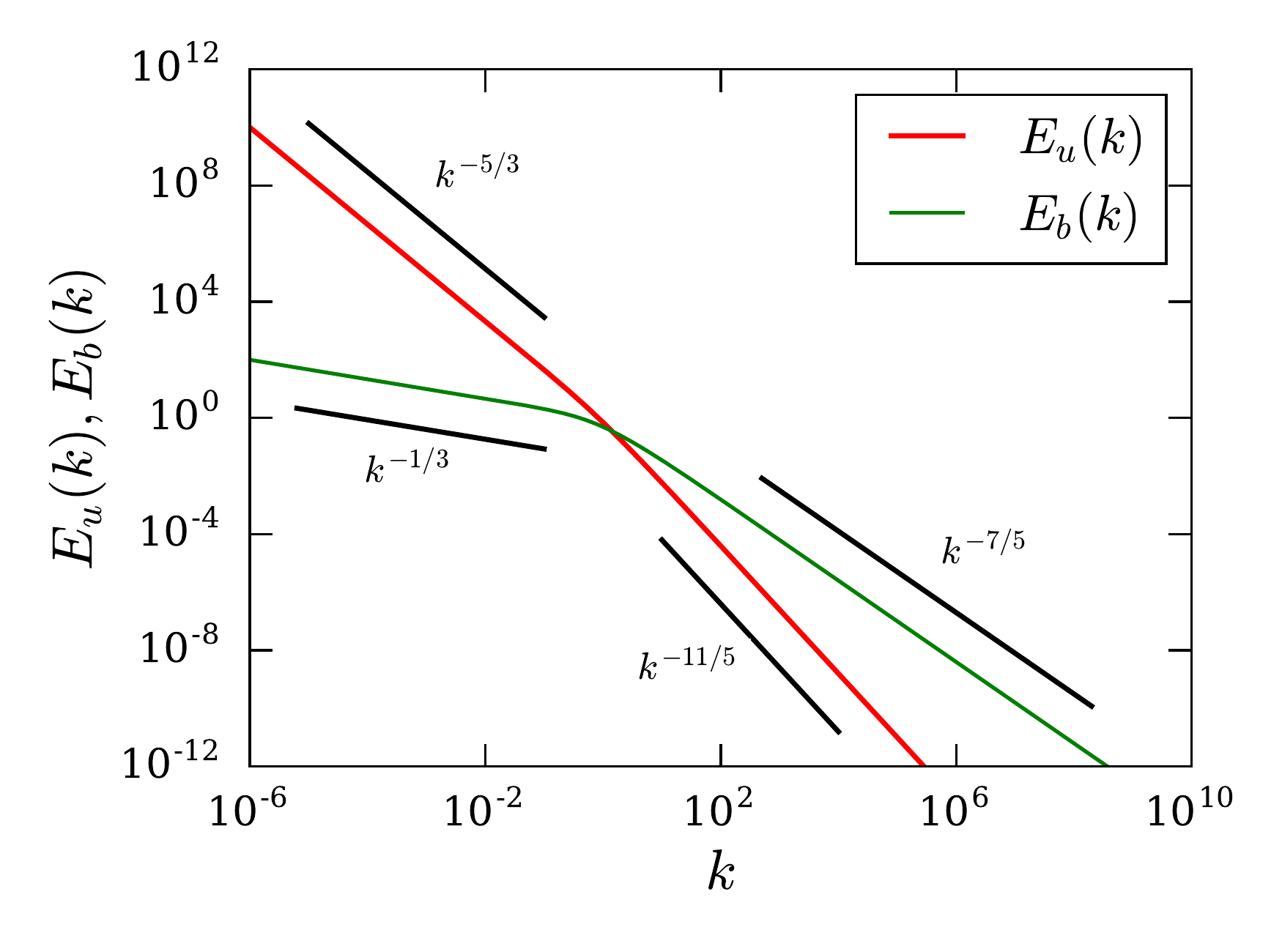}\\
         (a) & (b)
 \end{tabular}
\end{center}
\caption{(Color online) Fluxes and energy spectra for $N = 1.0$ and the total energy flux $\Pi = 1.0$. (a) Kinetic energy flux ($\Pi_u(k)$) is plotted in red and potential energy flux ($\Pi_{b}(k)$) is  plotted in green. (b) Kinetic energy spectrum ($E_u(k)$) is  plotted in red and potential energy spectrum ($E_{b}(k)$) is  plotted in green. In both figures,  black lines represent asymptotic behaviours in the extreme limits.}
\label{fig:SST_Flux_Spectrum}
\end{figure}

We numerically solve (\ref{eq:u_k_fifth_order})  using fsolve function of SciPy library in Python, which uses Powell's hybrid method to find zeros of non-linear functions. We choose $N = 1.0$, and  the total energy flux $\Pi = 1.0$, which is also equal to the total dissipation rate $\epsilon$.  We vary $k$ from $10^{-6}$ to $10^{10}$ in logarithmic scale.  These parameters can be treated as nondimensional with time period of large scale gravitational wave as the time scale, system size as the length scale, and large scale velocity as the velocity scale.  Using the numerically evaluated $u_k$ and $b_k$ we evaluate $E_u(k) = u_k^2/k$, $E_b(k) = b_k^2/k$, $\Pi_u(k)= k u_k^3$, and $\Pi_b(k) = k u_k b_k^2$.  The quantities are plotted in figure \ref{fig:SST_Flux_Spectrum}.

Figure \ref{fig:SST_Flux_Spectrum} exhibits  the fluxes and spectra of the kinetic and potential energies. For $1< k < 10^{10}$, $\Pi_b \approx 1$, $\Pi_u(k) \sim k^{-4/5}$, $E_u \sim k^{-11/5}$, and $E_b \sim k^{-7/5}$, which are the predictions of Bolgiano-Obukhov phenomenology for $ k < k_B$.  Surprisingly, there is no crossover to $k^{-5/3}$ scaling of passive scalar turbulence.  This is because $u_k \ll b_k$, hence $u_k$ cannot induce a constant kinetic energy flux.  We will show a more rigorous derivation in the next subsection.

Interestingly, for $k \ll 1$, we obtain $\Pi_u \approx 1$, $\Pi_b \sim k^{4/3}$, $E_u \sim k^{-5/3}$ and $E_b \sim k^{-1/3}$.  That is, $u_k$ dominates $b_k$ at small $k$'s that leads to Kolmogorov's scaling for the velocity field.  Note however that  $k = 1$ corresponds to $1/L$.  Hence, $k \ll 1 $ is possible in SST when the transverse length scale is much larger than the vertical scale ($L$).

In the next two subsections we will perform asymptotic analysis of (\ref{eq:flux_sum2}).  
 
\begin{figure}
\begin{center}
\includegraphics[scale = 0.6]{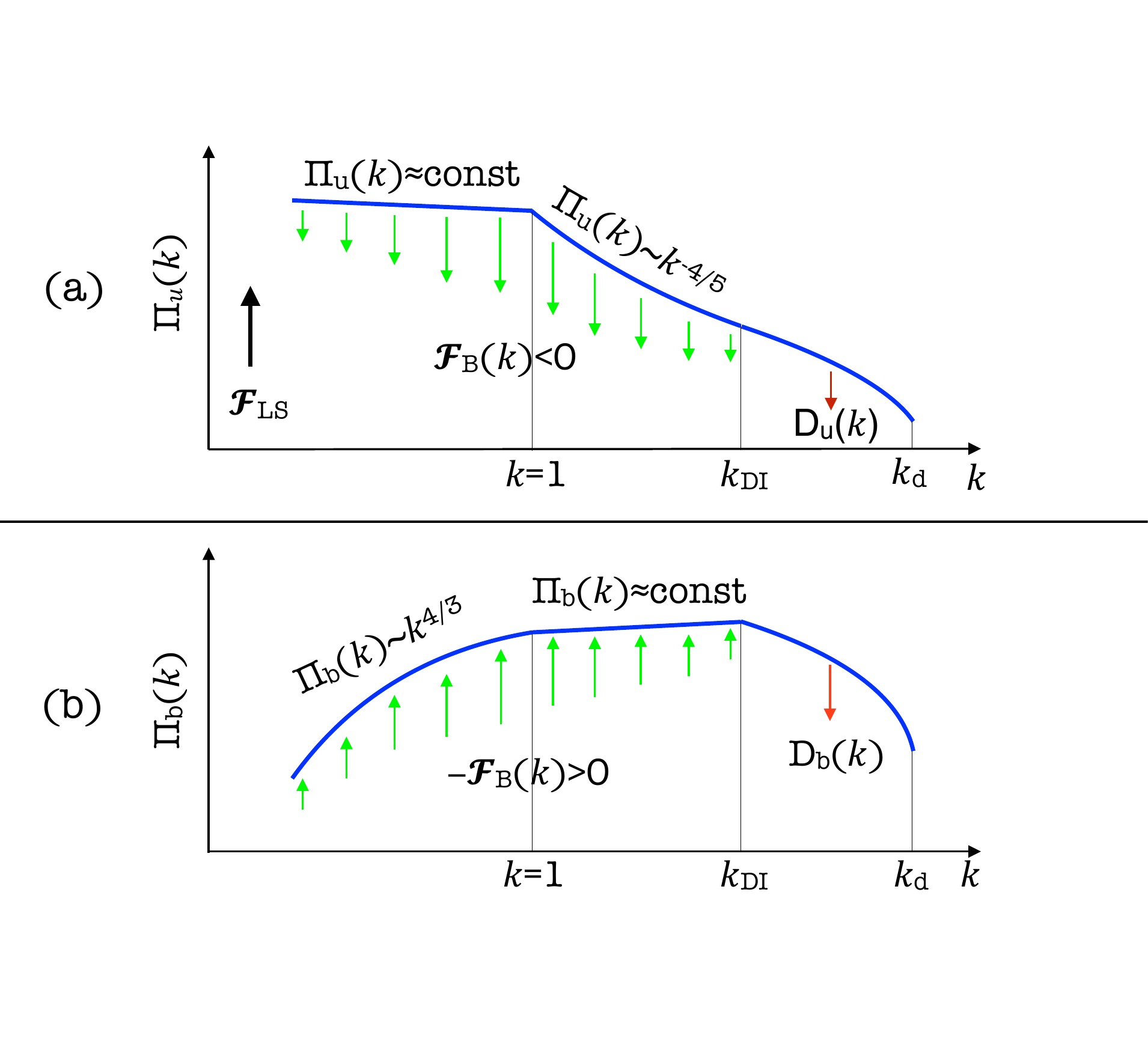}
\end{center}
\vspace*{-35pt}
\caption{(Color online) Schematic diagram of  moderately stably stratified turbulence according to the \emph{revised} Bolgiano-Obukhov phenomenology, which is expected in numerical simulations.  (a) Kinetic energy flux and (b) potential energy flux. Energy feed rate by buoyancy ($\mathcal{F}_B(k)$) is shown by green arrows. With $k \lesssim 1$, $\Pi_u(k) \approx $ constant, $\Pi_b(k)$ and $\mathcal{F}_B(k)$ increase with $k$ as $\sim k^{4/3}$ and $\sim k^{1/3}$ respectively. With $k  \gtrsim 1$, $\Pi_b(k) \approx $ constant, $\Pi_u(k)$ and $\mathcal{F}_B(k)$ decrease  with $k$ as $ \sim k^{-4/5}$ and $\sim k^{-9/5}$ respectively. }
\label{fig:Pi_schematic_sst}
\end{figure}

\subsection{Asymptotic analysis}
\label{sec:asymptotic analysis}

We examine  the dominant balance for the two extreme limits of (\ref{eq:u_k_fifth_order}).

\subsubsection{Case 1: moderately stably stratified turbulence for $k \gg 1$}
In this situation, $\Pi_u  \ll \Pi_b $, hence the balance is between $\Pi_b$ and $\epsilon$:
\be
\frac{k^3 u_k^5}{N^2} \approx \epsilon \Longrightarrow
 u_k \approx \epsilon^{1/5}N^{2/5}k^{-3/5}.
\label{eq:u_solution1}
\ee
Using (\ref{eq:force_balance_b}), $b_k$ is found to be  
\be
b_k \approx \epsilon^{2/5}N^{-1/5}k^{-1/5}.
\label{eq:b_solution1}
\ee
Therefore the kinetic and potential energy spectra and fluxes, as well as the energy feed by buoyancy  are given below:
\begin{subequations}
\bea
\Pi_u(k) & \approx & \epsilon^{3/5} N^{6/5} k^{-4/5},   \\
\Pi_b(k) & \approx & \epsilon, \\
\mathcal{F}_B(k) & = & \frac{\partial }{\partial k} \Pi_u(k) \approx -\frac{4}{5}\epsilon^{3/5} N^{6/5}k^{-9/5}, \\
E_u(k) & \approx & \epsilon^{2/5}  N^{4/5} k^{-11/5},  \\
E_b(k) & \approx & \epsilon^{4/5}  N^{-2/5} k^{-7/5}.  
\eea
\end{subequations}

Note that $u_k \sim k^{-3/5}$ decreases faster than $b_k \sim k^{-1/5}$. Therefore, buoyancy is strong enough so as to yield $E_u(k) \sim k^{-11/5}$ for the whole of inertial range, without a transition to   $E_u(k) \sim k^{-5/3}$  regime. Note that dissipation range starts after the inertial range.

A more quantitive condition for the absence of the second regime ($k_B$ to $k_\mathrm{DI}$ of  figure \ref{fig:Pi_schematic_sst_bo}) is obtained as follows.  Clearly, the Bolgiano wavenumber should be much smaller than the Kolmogorov wavenumber, $k_d$, which leads to
\be
N^6 \epsilon_b^3 \epsilon_u^{-5} \ll \epsilon_u \nu^{-3},
\ee
or
\be
N^2 \nu \ll \epsilon_u^{2} \epsilon_b^{-1}
\ee
In the above equation, substitution of the following expressions for the Richardson number and thermal dissipation based on the r.m.s. quantities~\citep{Verma_BDF_2018}:
\bea
Ri &=& \frac{N b_\mathrm{rms} L}{ U^2} ,
 \label{eq:Ri2}      \\
\epsilon_b &=& \frac{U b_\mathrm{rms}^2}{L}
 \label{eq:epsilonb}
\eea
yields
 \be
 \epsilon_u \gg \frac{Ri}{\sqrt{Re}}  \frac{U^3}{L},
 \label{eq:cond_epsu}
 \ee
 where $L$ is the length scale of the system.   Using $Ri \approx Fr^{-2}$, we obtain
 \be
 Re_b = Re Fr^2 \gg \frac{U^3/L}{\epsilon_u},
  \label{eq:Reb}
 \ee
  where $ Re_b$ is the buoyancy Reynolds number. As an example, for $Fr = 1.58$, \cite{Maffioli_2016} obtained $Re_b = 10430$  . Interestingly,  (\ref{eq:Reb}) is similar to that obtained by \citet{Brethouwer:JFM2007} for strongly stratified  turbulence.

 Since $\epsilon_u \ll \epsilon_b$, the above condition may be very difficult to achieve in  numerical simulations.  If we assume that $\epsilon_u = 10^{-3} \epsilon_b \approx 10^{-3} U^3/L$, for $Fr = 1$, Eq.~(\ref{eq:Reb}) predicts that $Re \gg 10^3$.  Such a flow would be difficult to simulate.  Therefore, we claim that the second regime of BO scaling is very difficult to find in numerical simulations. 
   It would be interesting to attempt to find this regime in a shell model~\citep{kumar_2015} or in some experiment.

\subsubsection{Case 2: moderately stably stratified turbulence for lower wavenumbers $(k \ll 1)$}

Equations (\ref{eq:u_solution1})--(\ref{eq:b_solution1}) indicate that $u_k \approx b_k$ near $k=1$.  For $k \ll 1$,   $\Pi_u  \gg \Pi_b $ implying that the dominant balance has to be between $\Pi_u$ and $\epsilon$:
\be
k u_k^3 \approx \epsilon \Longrightarrow u_k \approx \epsilon^{1/3} k^{-1/3}, 
\label{eq:u_solution2}
\ee
Using (\ref{eq:force_balance_b}), $b_k$ is found to be 
\be
b_k \approx \epsilon^{2/3} N^{-1}k^{1/3}.
\label{eq:b_solution2}
\ee
With the above $u_k$ and $b_k$, the evaluated energy spectra and fluxes, as well as the energy feed by buoyancy  in this situation are given below:
\begin{subequations}
\bea
\Pi_u(k) & \approx & \epsilon,   \\
\Pi_b(k) & \approx & \epsilon^{5/3} N^{-2}k^{4/3}, \\
\mathcal{F}_B(k) & = & -\frac{\partial }{\partial k}  \Pi_b(k) \approx -\frac{4}{3}\epsilon^{5/3} N^{-2}k^{1/3}\\
E_u(k) & \approx & \epsilon^{2/5}k^{-5/3},  \\
E_b(k) & \approx & \epsilon^{4/3}  N^{-2} k^{-1/3}.
\eea 
\end{subequations}
However, it is not certain whether the above scaling can be observed in realistic systems. The range $k \ll 1$ is possible in a large aspect ratio box,  but such systems could exhibit two-dimensional or quasi-two-dimensional turbulence for which ~(\ref{eq:flux_sum2}) is not valid.  Hence this prediction needs to be tested thoroughly in future. Schematic diagrams exhibiting kinetic and potential energy fluxes based on the revised Bolgiano-Obukhov phenomenology are shown in figure~\ref{fig:Pi_schematic_sst}.

\section{Conclusions} \label{sec:conclusion}
In this  paper, we revisit the celebrated Bolgiano--Obukhov (BO) phenomenology for stably stratified turbulence under moderate  stratification.  BO phenomenology predicts  a dual scaling for the energy spectra: $E_u(k) \sim k^{-11/5}$ for  $k<k_B$, and as $\sim k^{-5/3}$ for $k>k_B$, where $k_B$ is the Bolgiano wavenumber.  The  potential energy varies as  $\sim k^{-7/5}$ and $\sim k^{-5/3}$ respectively in the respective regimes. The transition to $k^{-5/3}$ scaling is based on the argument that the energy supply rate from buoyancy becomes negligible when $k$ is large, thus making density a passive scalar (such passive scalar behavior of density is observed in weakly stratified turbulence). 

In the present paper, we start with the constancy of total energy flux that yields a fifth order algebraic equation for $u_k$.  Numerical solution of the above equation yields $E_u(k) \sim  k^{-11/5}$ and $\Pi_u(k) \sim k^{-4/5}$, with no transition to the Kolmogorov-like scaling for larger wavenumbers.   The reason behind  the absence of the second scaling is that $u_k$ is too weak at large wavenumbers to be able to start a constant energy cascade.  The above scaling is also substantiated using asymptotic analysis.

In addition, we also derive the quantitative condition for obtaining the Kolmogorov scaling; it is given by $k_B \ll k_d$, where $k_d$ is the Kolmogorov's wavenumber.  This condition yields $\epsilon_u \gg (Ri/\sqrt{Re}) (U^3/d)$, which may be  difficult to satisfy in numerical simulations considering the fact that $\epsilon_u \ll \epsilon_b$.  However, it may be possible that such extreme condition for observing the second regime of BO scaling be satisfied in some shell models of stably stratified turbulence.

In conclusion, we  believe that our revised scaling of the Bolgiano-Obukhov formalism for moderately stable stratification will have important consequences in the modelling of buoyancy-driven flows.

\section*{Acknowledgement} \label{sec:Acknowledgement}
We thank Shashwat Bhattacharya for his valuable suggestions. A.G. and M.K.V. thanks PLANEX/PHY/2015239, and  A.G. thanks SERB Early career research award ECR/2016/001493 for funding support. 

\bibliographystyle{jfm}
\bibliography{bib/journal,bib/book}

\end{document}